**22**

# Efficiency of Biometric integration with Salt Value at an Enterprise Level and Data Centres

Bhargav. Balakrishnan
*Sutherland Global Services*
*India*

## 1. Introduction

Biometric have been an effective tool in providing the authentication for the authorized user to access the resources of an organization. They have been widely used in data centres and at enterprise organizations as it require lot of security; those are termed as information security (confidentiality of data). Even then how the hackers are able to trace the network and break the passwords. Is there any weakness in the design of the operating system? Why the designer of the operating system have not come up with any tool that can provide better security for the servers. As we all know that securities are applied at different stages of an OS like at system boot, before login screen and the final password checkpoint at the logon screen. The network is designed in such a way that each stage from firewall till user web access is monitored, then how the hackers are able to trace the flaw. To avoid this happening especially loss of data can be prevented by including biometric at higher level of security. Biometric is one of the tools that provides authentication only for the registered/ authorized users of that respective server. Once if it joins with SALT value (randomly generated value of any length) which is nothing but the password of the authorized user and maps with the encrypted value to authorize the user access on to the server, the server level security goes high. Biometric have not been interfaced with SALT value yet and used for authentication of authorized user's at server level. Whenever the security is applied on the server level especially for Microsoft Servers, the complexity of the password alone is not sufficient as there has been lot of possible ways designed by the hackers to break that password. Here the biometric when included will not allow the hacker to penetrate as it (Biometric image) is unique for every user. There will be lot of FAQ's regarding this type of methodology for user authentication at server level.

When the authorized user gets hurt in his finger for example, how the server can be accessed?

Solution Here the application should be designed in such a way that it accepts maximum of two thumb impression. When it goes beyond this the user has to log on emergency mode in server by pressing f8, which can be accessed by means of a complex password with minimal access to applications. (Will be explained in depth in coming topics)

How does biometric image and password maps with Encrypted value store in the NTDS file of windows 2003 server?

Solution: -The authentication pattern is similar to any authentication methodology that is followed in NT authentication, mail servers etc... A slight modification will include a biometric image + SALT Value that automatically generate an encrypted value which maps with stored encrypted value. The encryption algorithm should be changed on regular basis accordingly the encrypted value corresponding to each user will change.

Here how the biometric is going to help?

Based on the image the value is generated even though SALT Value (here it is user's complex password) is known to stranger the respective system of a user can not logged with his (authorized user) thumb impression, that's where biometric provides security at enterprise level or data centre.

## 2. Biometric techniques

There are different biometric techniques and some of the commonly known techniques are as follows 1. Finger Print Technology is an impression of the friction ridges of all or any part of the finger. A friction ridge is a raised portion of the on the palmar (palm) or digits (fingers and toes) or plantar (sole) skin, consisting of one or more connected ridge units of friction ridge skin. These ridges are sometimes known as "dermal ridges" or "dermal" 2. Face Recognition Technology is an application of computer for automatically identifying or verifying a person from a digital image or a video frame from a video source. It is the most natural means of biometric identification. Facial recognition technologies have recently developed into two areas and they are Facial metric and Eigen faces 3. IRIS Technology uses the iris of the eye which is colored area that surrounds the pupil. Iris patterns are unique and are obtained through video based image acquisition system. 4. Hand Geometry Technology include the estimation of length, width, thickness and surface area of the hand. Various method are used to measure the hands- Mechanical or optical principle 5. Retina Geometry Technology is based on the blood vessel pattern in the retina of the eye as the blood vessels at the back of the eye have a unique pattern, from eye to eye and person to person Retina is not directly visible and so a coherent infrared light source is necessary to illuminate the retina. The infrared energy is absorbed faster by blood vessels in the retina than by the surrounding tissue. The image of the retina blood vessel pattern is then analyzed 6. Speaker Recognition Technique focuses on the vocal characteristics that produce speech and not on the sound or the pronunciation of speech itself. The vocal characteristics depend on the dimensions of the vocal tract, mouth, nasal cavities and the other speech processing mechanism of the human body. It doesn't require any special and expensive hardware. The signature dynamics recognition is based on the dynamics of making the signature, rather than a direct comparison of the signature itself afterwards. The dynamics is measured as a means of the pressure, direction, acceleration and the length of the strokes, dynamics number of strokes and their duration. There are a lot of other biometric techniques like palm print, hand vein, DNA, thermal imaging, ear shape, body odour, keystrokes dynamics, fingernail bed. But these techniques are not been widely used in the authentication of the a person in attendance marking, server level authentication, authentication of a resident card holder as there are not feasible as the commonly used techniques which has been described above. As the authentication techniques should be feasible enough both in security and usability of the device. Based upon which only, the organization will accept for the implementation of Biometric authentication technique for their security purpose.

## 3. Evaluation on various Biometric techniques

### 3.1 False Accept Rate (FAR) and False Match Rate (MAR)

The probability that the system incorrectly declares a successful match between the input pattern and a non matching pattern in the database is measured by the percent of invalid matches. These systems are critical since they are commonly used to forbid certain actions by disallowed people.

### 3.2 False Reject Rate (FRR) or False Non-Match Rate (FNMR)

The probability that the system incorrectly declares failure of match between the input pattern and the matching template in the database is measured by the percent of valid inputs being rejected. This happens in some of the biometric authentication technique as it will give a negative result when the log is generated as the image it has authenticated is different which will be considered as a negative parameter.

### 3.3 Relative Operating Characteristic (ROC)

In general, the matching algorithm performs a decision using some parameters (e.g. a threshold). In biometric systems the FAR and FRR can typically be traded off against each other by changing those parameters. The ROC plot is obtained by graphing the values of FAR and FRR, changing the variables implicitly. A common variation is the Detection Error Trade-off (DET), which is obtained using normal deviate scales on both axes. This more linear graph illuminates the differences for higher performances (rarer errors).

### 3.4 Equal Error Rate (EER)

The rates at which both accept and reject errors are equal.ROC or DET plotting is used because how FAR and FRR can be changed, is shown clearly. When quick comparison of two systems is required, the ERR is commonly used. Obtained from the ROC plot by taking the point where FAR and FRR have the same value. The lower the EER, the more accurate the system is considered to be.

### 3.5 Failure to Enrol Rate (FTE or FER)

The percentage of data input is considered invalid and fails to input into the system. Failure to enroll happens when the data obtained by the sensor are considered invalid or of poor quality.

### 3.6 Failure to Capture Rate (FTC)

Within automatic systems, the probability that the system fails to detect a biometric characteristic when presented correctly is generally treated as FTC.

### 3.7 Template Capacity

It is defined as the maximum number of sets of data which can be input in to the system.

## 4. Basic setup of enterprise level security

As we see from the above diagram the security that is applied at each stage of a network. Even after applying these securities how the hackers are able to penetrate through the network and able to steal the confidential data's of many user's. When the user is accessing his bank account through net banking or when he trying to do a transaction of money over a network all that is required is security for his password and his account. Even then the hackers are able to get the username but getting a password is what his challenge is with which he can manipulate anything on the customer's account. A lot of these things are happening in today's present scenario. But how to secure these kinds of flaws both at a server level as well as at a user level is what is going to be discussed in depth in this chapter and the methodology that is going to be used to prevent this using the biometric and salt value along with the encryption algorithm. The biometric can't be used at a wide level at a Net banking as every user will not have a laptop or can't get biometric devices separately. In order to apply even that at an enterprise server level, ho to do that is what is going to be discussed in this methodology of server and application authentication at an enterprise level.

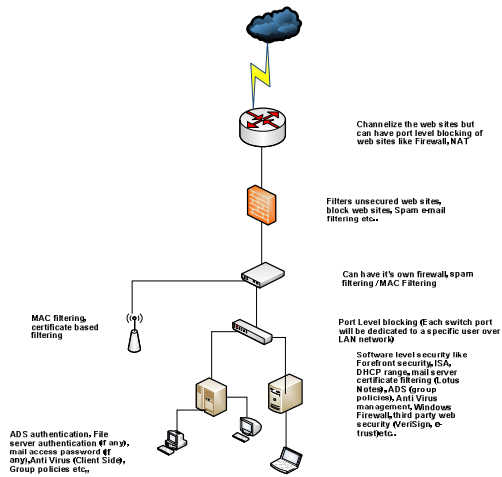

Fig 1 Security applied at each stage of a network

Each stage has its own encryption algorithm but having something included unique within an Encryption algorithm is what to make the data centres to have their information's keep even more secured. Each application has an encryption algorithm right from Cisco routers but they are also broken in many ways them the hackers are able to get the IP address of the internal network by some means. Even some organizations allow users at higher level executive to use Pen Drives on their official computers. If the antivirus installed on the computer is not effective then the virus /spam that has affected the other computer can penetrate into the network and can affect many other computers over the network. So what happened to the security of the information's that are stored on the server? The main drawback comes here is the server authentication are maintained with just the password and the encryption that comes with the server OS alone. But even though the password is set complex it is easy for hackers to reset the password. There are lot of encryption algorithm in today's world which are making the process of breaking up the security password. Even after using a lot of network monitoring many organizations are facing this issue. How to resolve these kinds of issues at Servers at enterprise level is the place the biometric and the salt value is going to play a vital role. As the biometric images are unique as we all know and can also provide a better level of security with both the SALT value and the encryption algorithm.

## 4. 1 Parameters of Biometric technique at security level

### 4.1.1 Permutation and combination

Why we have to choose permutation and combination while applying biometric at enterprise level? The main reason is to have a redundancy when there is a user gets hurt then what will be an alternate option. When we take the eye the possibility of generating a biometric image from a person will be two and it mainly depends on the characteristic of the light behind it that is the brightness. When there is some slight variation in the light that is generating this image can cause the authorized users from accessing and the probability that can be tried in this approach is also less. There are certain concepts like Voice, finger print where the probability becomes wider. The other techniques are also effective but each as specific criteria to bring that at complete enterprise will violate the security norm as well as it will be taken into account that is what we call as "Risk management". The live server are always are handled with a lot of risk and security measure taken for it will high. Let me explain you how this permutation combination concept is going to work.

For example, if I am going to be an authorized network engineer at an organization and have been given the permission to change certain things on the server regarding to the network and it's security monitoring. I have generated a biometric image with my fingers say 2 fingers from the left hand and 2 from the right hand. Now on that I have got fracture in my right hand. So the possibility of generating biometric images using the two fingers is there in the left hand. So the combinations that are accepted by the system is high and it becomes flexible for the authorized user to operate on the server and also secured as the images are unique to each person. Only the registered users along with their password (SALT value) and encryption algorithm that is getting generated internally once after accepting the biometric image and password of a user is going to map with the encryption table. The encryption algorithm can be varied on a weekly basis to ensure that the encrypted value are manipulated periodically to ensure high level of security at the server level as that is like the heart of an organization and the stage above it are like a wall or barriers for the hackers. Certain server can have an authentication from couple of biometric generated by the same person which will converted into the respective formats using any mathematical approaches and it is going to be discussed in topic where the generation of encryption is going to be done.

$$X1/X2 + SALT\ VALUE\ (Y1) = W1+Z1 = Final\ encrypted\ value\ (E1) \qquad (1)$$

Here X1/X2 are the biometric images in which either of them can be used. But in which biometric is this combination are more is in very less techniques. Then the ones that are having more probability will be the finger print and the voice. But the voice also has a specific drawback.. When I generate a voice encryption the application should filter the unwanted sounds that come apart from the voice of the authorized user then the probability of using the voice in authentication techniques will be high. As the voice is having a lot of combination like the finger print and can be converted into different format before it comes with a different format of password (SALT value) thereby providing a highly security approach of security. The main thing that should be joined with voice is the filtering of the unwanted sound from the background every time right from the registration of authorized user on a server at enterprise level. There are a lot of combinations that needs to be taken into account as the server needs to be accessed regularly so the technique that can process easier will be voice and finger print. Let us look into the other techniques and according to the priotization, reliability, usability and feasibility the biometric techniques will be utilized but having a common will make the process of authentication easier. Let us see a brief description on the parameters that are going to play an important role in the implementation of sever authentication technique at an enterprise level using the biometric and SALT value as a source of generating the authentication code. Then the code is going to map with the encryption process for authenticating an authorized user.

### 4.1.1.1 Priotization

The servers at enterprise will be undergoing monitoring at regular intervals and accessing the servers for various purposes will be high. Certain servers will be accessed at specific intervals like data base server, web server, net banking, ATM servers etc... Like at the end of the day to generate the complete report on the transaction and they are accessed only by certain authorized users who are technical specialist on that application and also who can generate the end of the day report as the data's which are seem as highly confidential like user's account number, Pin numbers, account details which are normally kept highly secured for which this biometric approach of authentication will make it highly secured. In this sector the biometric authentication type should be highly secured and feasible. So in these sectors the highly recommended approach with finger prints and then comes voice recognition. Why these approaches are feasible in this section? The main reason the probability of generating the finger print image is more than the other biometric methodology. When the authorized user needs to access, there is no requirement of other criteria's like brightness of the room, the voice filtering, the position etc. The finger print is quite a simple approach of biometric and also gives high security for the authentication. The best example will be the Yahoo Mail where Yahoo has got finger print approach for accessing the e-mails. The other methodologies of biometric generation are also having the advantage over authentication but it is the division where we use them. The biometric image once generated should also be stored secured and then the Salt value generation should be random. Every day the SALT value should be different and it should get updated to the authorized user. The device that is used is generate the SALT Value every 60 seconds are been manufactured by EMC2. The current models of this SALT value device are RSA SecurID 900, RSA SecurID 700, RSA SecurID 800, RSA SecurID 200, and RSA SecurID 520. These are some of the device that are being widely used in today's enterprise where the security is give the most important priority when compared to other parameters of an organization policies.

### 4.1.1.2  Reliability on Biometric Techniques

Biometric is highly reliable when it comes to information security. How it is going to be a feasible approach when it comes to authentication at enterprise level? What are the things that needs to be customized in server OS especially Windows, Solaris? As customizing the server should be after getting the necessary approval from the OS developer and the license provider that is Microsoft/Sun whichever is going to be customized according to these authentication requirements. When the necessary approvals are processed and this customized OS needs to be approved by security norms designing bodies like ISO that this approach of server authentication can be practise. Once this is accepted then the methodology can be widely used in the enterprise level. So we are going to see the areas of justifying this methodology that is going to tell "HOW RELIABLE IS THIS METHODOLOGY?" 1. When this approach is implemented the possibility for the hacker to steal information becomes less as both the Biometric value and the SALT Value is going to be unique. Once these two numbers are going to joined as 0′1 and 1′s using any calculation like XorY, Xand Y, XnorY etc…Then the number formatting is changed and then when encrypted using an encryption algorithm the output will be completely and doesn't gives even a clue on what number or image is used. Even if the number is able to decrypted getting the same biometric image is hardly possible 2.The risk that is involved in maintaining these biometric image are high but there is a modification that is done for avoiding this risk and in a secured approach which will discussed in depth in the topic that is going to give a complete explanation of this biometric technique 3. The reliability on this biometric approach of authenticating server access can be high as the Biometric technique that is chosen is based on the maximum combination not with the least combination where it can be of a risk and the management won't agree for that approach. All that management requires from its point of view is an application that can be feasible and at the same time keep the information's of the organization and the client safe and secured. This methodology will be highly as it is going to be an integration of know approach but encrypted and combined in a different approach which has the capability of getting compatible at server level more easily with high reliability 4. The finger print that has been know for many centuries as a source of authentication but it was ink based that is pressed on a paper by an individual when there is an election to avoid misusing the policies of making another vote by the same candidate. So this gave a unique approach which was slowly being used for the authentication purpose in the country's visa card to authenticate a resident expatriate. So the finger print has been widely used in various application and sectors. That's has been reliable and feasible in authenticating a user 5. The finger print methodology has requires simple enhancement over the existing keyboard. The keyboard needs to be interface with the finger print reader which should have a driver that should be getting installed automatically when it is interfaced with the port on the server. That should have an application that should have a transfer the image to the application that should to in turn go through the entire process of authentication which will be explained in the section where we are going to discuss on the complete process involved in this authentication 6. The finger print is a technique that can be used in this scenario. Let us analyze on other techniques also and bring a complete analysis work on that area of work. This will be a valid justification on the exact priority of the biometric techniques 7. The biometric techniques are more reliable when compared to third-party software requesting to remember an image as a source of authentication for the users over the Net Banking. This needs to be a part of risk when the management point of view especially when this kind of methodology is applied at Net Banking concepts 8. Now the bank ATM card systems are slowly bringing up this technology apart from the PIN system as it will make the process of authentication much easier and with high security.

### 4.1.1.3 Usability of the Biometric techniques

When it comes to the usability of the biometric devices, it has been simple as the installation is done by the infrastructure team along with the maintenance. The biometric has been really user friendly in terms of registering their image like finger, eye, palm etc... But when you to login using it there has been a fault tolerant that sometime if the brightness of the eye was not equivalent to the brightness that was there during registering it might not accept and this sometime makes the user to use the password to login into the computer. But when it is going to be designed at a server level it should not be the considered as a negative parameter. Here the biometric that is going to be should be highly advanced in authenticating the user in a much quicker way. It should try to filter the brightness and auto adjust itself so that it should only take the exact picture of the authorized and not the brightness behind the picture. Apart from the finger print other biometric techniques needs filtering. This will increase the options of using the finger print in the biometric techniques. Transaction authorization: A subject can execute a transaction only if the transaction is authorized for the subject's active role. With rules 1 and 2, this rule ensures that users can execute only transactions for which they are authorized. Here the sensor is the device that is going to identify a authorized user's biometric image. When a user comes before it or swipes through the device it will take the image then it will it will go to pre process, the image is will be converted to the required format as designed in the parameters of the features followed by generate the template (Generate biometric template customized based on feature parameter). In the pre-processing, it has to remove artifacts from the sensor, to enhance the input (e.g. removing background noise), to use some kind of normalization, etc. Then it will be stored in the database of the Biometric device. Then again when the swipes, it will go through the process of customizing then goes to matcher and then check it matches with the one stored in the Biometric database. Here what type of conversion is being used? The algorithm that is used is Matching algorithm. The matching program will analyze the template with the input. This will then be output for any specified use or purpose (e.g. entrance in a restricted area). So the probability of this algorithm securing the biometric image when compared to the biometric image into 0's and 1's will be discussed in the later topics. As the image when converted to 0's and 1's either by binary, octal, hexadecimal. It is then applied to digital conversion like 4B/5B, 6B/8B format then converted to the number makes a rearrangement of bits and it will be of high security when applied at Enterprise level. This makes the complete process of the Biometric authentication process. This diagram will be common for all the biometric but the encryption algorithm and approach of Biometric authentication varies a bit. In some device it will take only a biometric image for authentication like laptop, resident card authentication. But when you take for entrance security it has biometric image with a key to authenticate a user for his attendance. But how effectively they are used is comparatively less as the users finds it tedious with the work pressure they have and this process is mostly ignored in many places. Normally they have password authentication for access the organization or card system which is swiped and mark the attendance. But how far the card system has been effective is very less when compared to Biometric authentication. Even in major bank it has not been implemented. The Biometric is not implemented at entrance, locker section and the server room where all the confidential data's are stored.

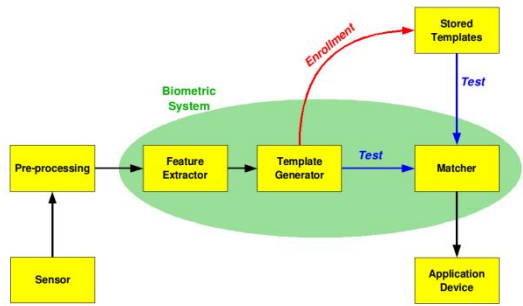

Fig 2 The basic block diagram of a biometric system

**4.1.1.4 Feasibility of the biometric authentication**

The application designed for the biometric authentication has been highly feasible as it is just to store and authenticate the authorized user when he accesses a security location. The authentication should have a proper a backup and restore system as to a source of redundancy if the device gets damaged or the image template gets corrupted. As there will be fault tolerance in any software as it doesn't have a specific reason for it to get corrupted. So that is the only that need to be really careful. As the authentication devices are highly feasible bu the damage of the device depends on the life factor that is quoted for that device. So the backup of the authentication template has to be taken on a regular basis along with the report logs of authentication which forms a part of security auditing. When the auditing is done for giving an organization with the ISO certification authentication process is a part of it. So this biometric authentication should be proper and it should be approved approach so that the organization can be secured in keeping their data's and client information with high security. The feasibility depend on many factors like change management, updation of firmware, risk management. The feasibility as we all know is classified as economical feasibility, technical feasibility and operational feasibility. When all these conditions are satisfied then only the using a particular biometric technique will be approved in a enterprise organization. When the biometric is used at enterprise it should be reliable, quick on authentication and price for the installation should be reliable. But when it comes to server authentication all matters is Information security which is of high priority than all those feasibility of biometric authentication. Time is not major constraint when compared to Security. The security level is considered and the analysis of that which will seen in the coming topics. The Information security have been the major constraint and for which security at the network and server level is increased periodically to ensure that the data when it is transmitted over the network are not been easily decrypted by the hackers. It is a real challenge for the people at the security domain and auditing vertical of security. As the hackers are working very hard to track and try to create a lot of problems. But how this biometric is going to help in this process is going to being discussed and also later works on integrating biometric with confidential data's during transmission of data over the network. This will ensure that the data's are safe in both Inter ad Intra network locations. This type of strategic approach is much needed for this security level. Without having a proper approach towards the parameters which are mentioned selection of a biometric migh go wrong. So follow designing by means of the above parameters.

## 5. Proposed Biometric techniques

This is the proposed biometric process flow for authentication at sever level mainly we call as "Enterprise level support" where huge data's of customers, client, companies confidential data's are stored. At this level generation of report log is mandatory which is going to be generated at the EOD. Along with the authentication at the network level should be monitored and should be generated that is going to form the consolidated report for the day. These complete consolidation of data's at the end of the year is going to presented for the auditing based upon which the security policy of the company can be seen, Many proposed model of security are available which is customized as per the companies and used for the generation of the security audit reports. Here at registration process the Biometric sensor is going to be getting the biometric image from the authorized user and then going to ask for the SALT value. Both of them are converted to the respective binary format, then going to perform the Logic gate operation which will rearrange the arrangement of bits.

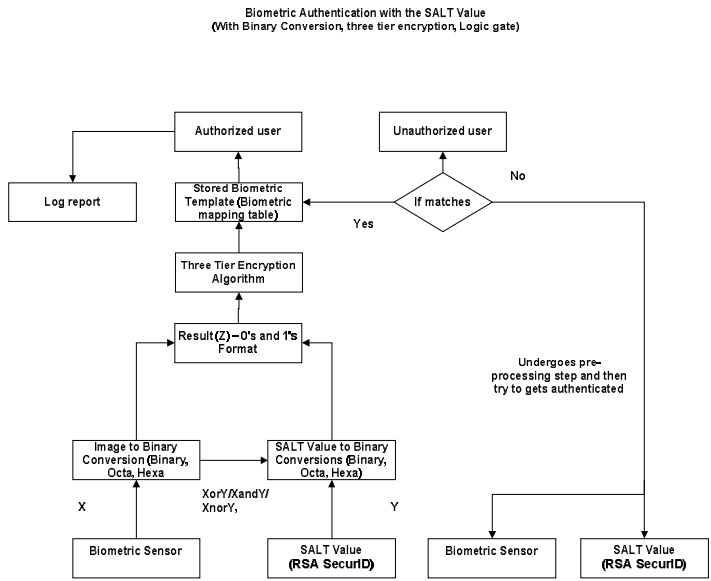

Fig 3 Biometric authentication with SALT Value for server level authentication

Then at the Three Tier encryption algorithm the encryption is going to happen which is then stored as the generated template of the biometric image. Once the template is generated when the user logs in based on the biometric image and the SALT value the calculation automatically map with the generated value. Which will then going to authenticate the user? Here comes the question how the saved template is going to authenticate when the saved SALT value keeps changing. The application is also simultaneously within the server as the same SALT value which is there with the authorized. So the encryption table will have generated template value updated accordingly and it will easily authenticate the user. Then where is the security. Here the entire uniqueness of accessing is the biometric image which cannot be easily generated. Apart from this SALT value will be in an encrypted format so tracing out the possible value will take many years for the hacker to break through. But they will never be able to manipulate any biometric image of an authorized user. That is the most important part of this methodology. Here the main intention was to prevent the unwanted access of servers by unauthorized users for checking the information's on the server without clear information. At the same the authorized user is not have to write the password and keep anywhere as it is his biometric image and SALT value that is randomly generated. The authorized will have to be careful with the SALT generating device. The administrator should be set with a lot of policies that for the security of the confidential data's. In the above you can see that the report log that needs to be generated it is something similar to the System log file which comes on the authorized who has accessed that particular server. The report log is generated for all the servers that are on the network of that organization. This report needs to be consolidated to get the final EOD report. It will give a clear picture of the user access control for the servers. If there is any loss of data's it is easy to trace with the authorized user log report. As there will be unmatched/ invalid user report that would be generated for the unauthorized user. This will make the process of tracking the unauthorized users simple. This is the main advantage behind this methodology. The IT infrastructure manager should be careful on the report log generation and should ensure that the LOG reports are generated at the EOD is properly consolidated for all the server and network infrastructure.

This methodology is going to be an enhancement in the current OS and it is going to be integrated with the necessary approvals from the OS developers like Microsoft, Linux, Solaris etc.. Once this approval is done then the customization of the OS can be done and the testing at the security level, feasibility level (mainly restoral level), reliability are going to studied in depth before executing at the Enterprise level. Here in this chapter this analysis is going to be seen in the coming topics. Why the format of 0's and 1's are being used and with a lot of rearrangement of bits. This is because when the hacker is having the ability for finding the number but the number format of 0's and 1's are quite difficult as it will be quite unique. So the hackers will not be able to identify that the exact that is being carried out with 0's and 1's. Let us see some of the analysis with the rearrangement of the bits and without it the security level will be comparatively less.

## 6. Generating Three Tier Encryption Algorithm

The steps of using this encryption methodology are as follows step 1. In the first step the RSA algorithm will be carried with the following modifications a. Consider two prime numbers as 11 and 13 b. N= P *Q i.e.   143 c. M = (P-1) (Q-1) i.e. 120 d. D is the decryption key Example 3 which is a prime number e. E = D inverse (mod n)   i.e. 47 f. Let the password be "Hello "take the ASCII value of the password covert it as 7269767679 g. Concatenate this ASCII value with a SALT value (Randomly generated number) say 34 i.e. 247172101086 h. Finally multiply this with the Encryption value to get final encrypted word 9886884043440 There are certain constrain which are modified and the requirement in RSA Algorithm are as follows  a. The minimum requirement for P and Q values in RSA is 2048 bits which gives the utmost security to the file that is being transferred b. Modification is inclusion of ASCII value conversion and SALT Value. Here SALT is being left user defined c. The P and Q values are also user defined that is also a modification d. At this you can use any encryption algorithms which are being updated. Step 2.The above arrived result through RSA – 9886884043440 will be converted into 0's and 1's using number conversion. The above encrypted data (9886884043440) will be converted as 10010110110111001010010101. This is for binary in the same way it can be done for octal /hexadecimal Step 3.This number conversion will be modified using Digital Encoding (Either Line or Block Encoding). Advantage: - Rearranges the bits of data i.e. 0's and 1's. Then use any of the line encoding schemes like NRZ, NRZ-I, RZ, biphase (Manchester, and differential Manchester), AMI and pseudo ternary, 2B /IQ, 8B/6T, and 4d –PAMS and MLTS that will convert the number which are being as binary in the above as follows Let us consider 4B/5B Block Encoding for the replacement of bits that were generated using the binary conversion. The output that would be generated by using the reverse conversion process with be different from the generated using the RSA algorithm. The output that is generated us mentioned below.
100110111011011110101011001010101101001110111111101001101110110111101010110010101 001101110110111101010110010101011010011101111111010011011101101111010101100101010 110100111011111110

| 4 bit value nibble | 5 bit value symbol | 4 bit value nibble | 5 bit value symbol |
|---|---|---|---|
| 0000 | 11110 | 1000 | 10010 |
| 0001 | 01001 | 1001 | 10011 |
| 0010 | 10100 | 1010 | 10110 |
| 0011 | 10101 | 1011 | 10111 |
| 0100 | 01010 | 1100 | 11010 |
| 0101 | 01011 | 1101 | 11011 |
| 0110 | 01110 | 1110 | 11100 |
| 0111 | 01111 | 1111 | 11101 |

Fig 4 4B/5B Substitution Block Encoding

Step 4.In this step the conversions of data into 4B/5B will be converted back into numbers using number conversions. This is reverse process of Step 2 Conversion back to binary will be give different encrypted word because of the usage of 4B/5B line encoding. The solution will be 10205099. Here also the conversion can be any one of the following binary/octal/hexadecimal. Step 5.In this step the above obtained number in step 4 10205099 will be considered as the X Value. This will be substituted in the Mathematical Series. Here in the example the sine series is being used. Formulae: -  $\sin(x) = X - X^3/3! + X^5/5! - \ldots$ Here X is the encrypted value 10205099. The series is used defined say N=3 then the series will be till $X^7/7!$ Then the final result will be 10205099 - 1771333826010.833333+ 246018586945945274.37 = 176887364014124447176.45856481478. Use the round off function to get the final encrypted word as 176887364014124447176.

**6.1. Advantage of using digital encoding, number conversion and mathematical series**

The main advantage of Step 2, 3, 4 is in Step 2 the encrypted data obtained by RSA is converted into 0's and 1's. Then by using Digital Encoding the rearrangement of Bit's are done. Finally in Step 4 the reverse process of number conversion. What it does? The hacker will never get a clue of this process that is being carried unless he gets an idea about this algorithm. Then Step 5 also a vital role as here the number X i.e. the value obtained from Step 4 has to be determined by the hacker, for which he should what is used, if found what mathematical series used which will takes ages to refine.

But for an organisation to encrypt and decrypt will be a simple as the process involved in each data encryption will be stored in their database. So this twist in the algorithm will be playing the most important in preventing the hacking of data's. How this methodology gives utmost security to the file at the same time increases the complexity in identifying the content by the intruder. These are being described below If the Intruder gets this encrypted word the following things are to be determined. Determining those values is a long process and finding those will take many years in order to arrive at the conclusion 1. The value of N i.e. the length of the series has to be determined 2. After finding N values the value of X has to be determined that has been substituted in the series 3. In the line encoding process the split up of the bits has to be determined like 4 bits, 8 bits and so on 4. After determining this, the type of encoding has to be determined and the substitution used as in the B8SZ where 8 bit value is substituted in place of continuous 8 zero's 5. Based upon which the entire two stages can be revealed from this the first stage can be proceeded that is RSA instead of that AES, SHA, MD5 any encryption algorithm can be used 6. The speciality of RSA is in determining the prime numbers P and Q which itself will take many years to determine.

The end user can be a data center, search engine etc which will get utmost security because of the usage of Line Encoding and Mathematical series. The line coding will convert the original encrypted word into duplicate encrypted word by using the following) i) binary/octal/hexadecimal the encrypted word is converted as 0's and 1's ii) then line encodings is used. This will act as a protection. This will be even more protective by using the mathematical series. On the whole the methodology will be a secure path for the transfer of data's. Time for generating the Encrypted file using this method will be comparatively less in the high end PC's with dual core processor and above with 2GB RAM with processor speed of 2.2 GHz. The RSA encryption of about 2048 bits will take time other steps will take fraction of seconds for generating the desired output.

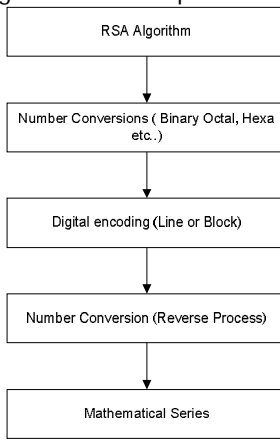

Fig 5 Diagrammatic representation of entire encryption process

This will give the complete idea on this encryption algorithm flow. Here the important step is in the replacement of bits as that is making the complete change in final encrypted result.

### 6.4 Advantage of this encryption methodology

There are various advantages of this encryption methodology which are as follows 1. In the file transfer preferably in the low privilege servers which are an endangered place of hackers 2. In the WAN where the data transfer is not that secured, in order to give a firm security this methodology can be adopted 3. This methodology will be of high value in the defence sector where security is given high preference. Using this methodology the hacker will not be able to trace the ideas unless or until he is well versed in the mathematical and electrical technique of disclosing the data 4. This will also play a vital role in other sectors like Bank, IT, Aero Space and many more where the data transfer is given more security. These are some of the advantage of this encryption in secured file transfer over the low privileged and it will be to secure the server at the same level of security.

## 7. Proposed server authentications (Complete Analysis)

This is the proposed authentication model which is going to be integrated with the current server level authentication procedure. This manipulation will be done with a lot of software testing as to avoid to any flaw in the live operation. Here the fault tolerance should be replaced by a redundancy procedure which is also discussed in this topic.
The biometric integration with SALT value is explained below
Step 1.E.g. Biometric Image -> Binary/Oct/hexadecimal

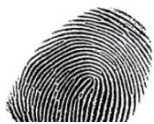 --------------→ 01000111100000111101110000011000010000001111000 – (1)

Step 2.SALT Value (Randomly generated value used as user password -> Binary/Oct/hexadecimal
Each user password will be joined with a SALT Value and then converted to respective format. SALT Value is generated once and given to the user. User needs to remember his password and SALT value which he will get the RSA secure ID device.

   cristopher2101 + (concatenating) 2341 -> (01000111000100001100) (0010101010111) – (2)

Step 3.Converted value of Biometric image and result of SALT value + user password

   01000111100000111101110000 (OR) 01000111000100 = 01110101010101111110010101 – (3)

Step 4.Then apply this output to the encryption process (Three Tier Encryption Algorithm) which will do the replacement the replacement of the bits and then the output will be a number of the format as shown below.

 01110101010101111110010101 -> 3242323131414113 -> 23456777888899989799712323234 – (4)

Note: - The above value is just an example value not the true value.

The conversion is done as per the above example. The conversion format can be varied as per the requirement but the steps involved in the conversion will be as per the above mentioned example. Once this is converted in the above mentioned format, the hacker will just see it as a number but to decrypt this value will take many years and then to generate the image will not help the hacker in any ways to penetrate into the server thereby stealing the data's. This replacement of bits is done along with image conversion and concatenation of SALT value + password is only to bring about confusion for the hacker in tracing the original value. The value obtained after conversion will no way provide a trace on what is used in the conversion process. To make an analysis on this is a difficult task as the following things needs to be analyzed. In the authentication process even decrypting the encryption algorithm will be of a big challenge even though the steps used seems similar but input that are unique and especially biometric image is unique as well as SALT value changes for every server login and it is simultaneously matching with the template with the mapping output generated simultaneously. So penetrating and making a change is highly impossible. But that is how the authentication should work at the enterprise level and there should be a proper server authentication procedure 1. No of bit used in conversion 2. The value joined in the process concatenation (Password + SALT Value) 3. The value of image (which will generate only with the authorized user) Eventhough hacker derives the step 2, for step 3 he needs the authorized user to access, which is no way possible. That is where biometric provides an effective security feature with encryption. This methodology of Encryption has been designed in such a way that the authentication process is secured as the time to authenticate is also less. In the step 4 the output that is shown is how the value appears after the rearrangement of bits and after applying the Mathematical series. So the complexity of the output will be very high and also make a trace of exact authentication flow will be quite difficult. That is going to be final template and end of day reports are going to be generated based upon this authentication flow. When the encryption is done all that matters it the time to take the input, generate the output and authenticate. So how this going to be calculated will be show with a breakage with time duration in each stage of authentication process. We will see the complete analysis for other authentication techniques and also see which is going to be effective in authentication, probability of generation, easy to generate a biometric image with being less affected with the environmental effect like Sound, brightness etc... Then we are also going to see how the biometric is going to be used in message authentication too. That is going tell the positives of Biometric usage in authentication procedure at the server level.  Let us know the exact manipulation that I have proposed for the redundancy in server level authentication when we use Biometric authentication. When the authorized has got hurt but has to make change in the biometric image to authenticate the server to Login when needed. How can we do that? Is that any procedure that can be done with high level of security and without breaking up the security norms of the organization and the client? This will be done with a proper approval from the management team of both the organization and the client. How is it going to be done is going to be seen in the next section of this topic. Here there are going to be two options that will be there in this application Update and reset but that can be seen only in the "emergency access mode". Here the access for the application will be very minimal as this mode is dedicated for the only the update or reset the biometric image by authorized with a specific password that is again generated using the RSASecure Id device.

This process is going to allow the authorized to go and change the biometric image in emergency or a periodic updation in the biometric image to make sure that the combination provided should periodically been changed and also make the authentication process go without any flaw. This is also used when the authorized is hurt. This can be done with a proper approval from the managers of IT, change management. IT security managers, risk managers etc…Let us now see on that process in depth and the procedure that needs to be followed before making those changes in the live servers.

### 7.1 Authentication at Server Level

Let us see how the redundancy in biometric image can be generated in emergency level that is when the administrator has met with an accident or due to some unavoidable circumstances. It is pretty much simple procedure but this is also highly secured methodology of accessing the server. 1. When we press F8, the OS opens in **Safe Mode** 2. In this another option needs to be included for server OS alone is **"Emergency Access Mode."** 3. When we access this, there will an option to insert **biometric image, generate new SALT Value and press update + reset button**. Only that window alone opens. This will not allow access to any other resource on the server. Let us see the advantage of this methodology.

In this portion of the OS this option needs to be brought about and then the same needs be linked with the application too which needs to go through some of the process of approvals in risk management, change management. Normally bringing that change is not an issue but this option is linking with the access control, application access control and its database where this biometric images get stored. The complete analysis procedure will be seen in the coming topics. How this procedure is going to be implemented is what is going to be seen and the time that is roughly required for the resting of this modification. So there is going to two modification (1) inclusion of Emergency access Mode (2) Integration of Biometric with user password at Server login authentication. How this option is integrated with the application that is installed inside which will be authenticating the user in place of server authentication which includes only password. So that is where it is going to be a real challenge for the developers who are going to make this change with testing, approvals etc. Let us see how this entire process flow for this modification is going to be made. Here there will be a doubt that why the reset of the new biometric image can't be like change of new password at the login page as the biometric image can be changed periodically the following the main security reasons behind not keeping that option there are as follows (1) It will become an option that would not be known to the unauthorized user to misuse it in the absence of the authorized user. This option which is integrated in the safe mode should not to known to anyone else other the authorized users of that server and the management executives. The integration is complex as the updation should happen properly when the biometric images are changed it should generated the final template and then when the user login back again it should be able to properly authenticate the user without any issues. Those are some of the places where the testing needs to be done and then deploy this OS in the live environment. Let us how the management going to take a decision on this change. The management which will be the main body for the approval of such important options like this which is going to be a part of the redundancy in the live operation. When a biometric image is going to changed or going to add a new biometric image. Here we can see how the management view a modification when it is brought about in an OS. Here we are seeing the parameters like % of validity, % of redundancy,% of probability, % of feasibility that are normally used to authorize a modification.

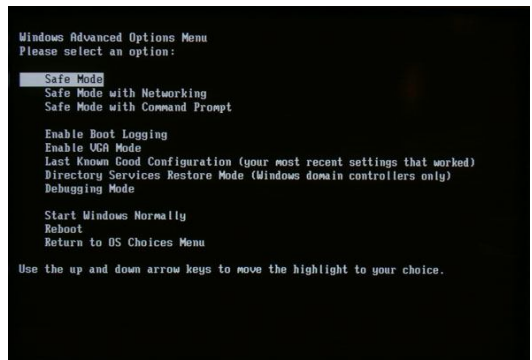

Fig 6 Current Safe mode options

When a change is brought about the modifications needs to be discussed with the above mentioned and justifies the reason why this change is brought about. How this change is going to help in the authentication level. Here it is all about the redundancy step followed in authenticate. The entire process flow should be explained and also tell them if a new biometric is inserted by an authorized how the updation happens and how that changes and the new authentication results can be seen in the report log. What are the key data's that needs to be seen a report, all these things needs to be explained to the management as they are the people who are also responsible if there is any loss of data by chance. How to trace an intruder's access from the log and also how to track his network path is what are the queries that a management will have. The justification should be from the development team, infrastructure team and IT security as those are bodies who are designing this application. What are the justifications for this modification? In this "Emergency Access Mode" the password is generated by a RSASecure ID which keeps changing the code every 60 seconds along with the password for the emergency access which will allow the authorized user to access the application with his server logon password and RSASecure ID code. Then the modification of the biometric image is done and saved in the encryption table. How this process is going to be secure approach for the modification. Here two things are unique (1) Code generated by RSASecure ID device (2) The user biometric image can't be caught and misused. Here when the biometric image is changed in the emergency situation or when there is a need to add a new profile how it is being done. When it is adding a new profile even the RSASecure ID device needs to be registered with this as the new user should be able to access the server without any issues. That is the reason once changed it needs to logged in and checked if that works without any issues. This needs to be carried out in the testing phase and not in the "live environment" as this will be a very costly issue if there is some unavoidable circumstances where the authorized is not able to produce his biometric authentication. Have 3-4 authentication techniques for a server authentication is always not advisable as it is like giving an option to the hacker to know the process that we are trying to manipulate. It should be a unique approach and there should be no trace of this modification to anyone even within the organization. In today's world the approaches are being leaked out in Media and the hackers are consolidating those techniques to hack some valuable information's from many data centers. How to avoid this is by maintaining privacy within the organization on certain information's that are related to the confidential data's, security techniques and policies behind it. Once if these things are known to any of the user within the organization they can try to misuse it using any third party tool. If the user is not given access to a information's. He will try to threaten the System admin and can try to manipulate the things within the organization. To avoid all these things the security policy and methodology should not shared to employee even though he is friend or relative of the System admin. That is the reason why the agreement needs to be signed by the system admin and organization as a agreement normally called as OLA in management term. This will allow the system admin to take a risk on this as it will in turn going to question him and not the management by the Client. So this will bring about a strict policy in the security approaches. This is how the process is secured even in the modification or adding of profile in Biometric authentication. Some of the advantages of this approach can be seen before seeing the complete advantage of Biometric authentication over current authentication methodology which are as follows (1) the NT authentication will be highly secured protecting the data's on the server (2) The possibility of breaking up the password will be highly impossible, as the encryption algorithm is changed on regular intervals along with biometric image on a quarterly basis (that is image of another finger of the user). The possibilities/ probability of changing the encrypted value is high with biometric and encryption (3) the approach for encryption is simple and decryption process for hacker is highly impossible. This methodology has lot of other advantages which will be seen after on how to frame this methodology on the basis of ITIL framework. Then we see on the analysis of each biometric technique on the basis of the parameters like error rate in authentication, error rate in initial registration, error rate in accepting new user, error rate in other factors like Light, sound etc. Finally in this chapter we are going to see some information on the security policies that needs to adopted by the organized for this methodology as it involves a lot of confidential information's before getting an authentication to a server and this methodology is specifically designed for the Enterprise level date centers. Finally but not the least we will see the future modification and other techniques going to brought about using the similar kind of methodologies. These are some of the things that are going to be discussed in the coming topics. As we have in the Fig 23 there is "NO" condition which tells there requires some modification so what can be the possible reason behind it will be (1) % of redundancy (2) % of error in wrong authentication acceptance (3) % of error is not accepting a authorized biometric image (4) % of flaw in application (5) % of time taken in authenticating at critical situations (6) %of feasibility in using that application are of the some of the common factors that comes in the minds of a managements. As a management employee he won't see on how this application is going to function but on how it going to keep the information secure as well as how it helps a n organization to drive a business easily with it secured approach in maintaining the data's of it clients. They see that whenever a organization needs to be retrieved from a specific server the authorized user should be able to get it without any issues in getting the authentication from the server. If the server is not accepting it then going to the Emergency access mode is quite a risky approach as it is the live server which he is turning down for a minute which is going to have a negative impact on a organization. The approach over here should be different and this is what going to come under the account of % of authentication acceptance.

% of acceptance = No of acceptance – No of rejection/ total no of logins – (1)

% of EAM usage = Total no of valid access – total no of invalid access/ Total of access in EAM – (2)

% of redundancy = total number of possible biometric image generated/Total possibility by that biometric technique * 100   - (3)

These are some of the parameters which are normally calculated at the end of the day report which will show the complete statistics of the biometric authentication results at the end of the day. Now this information's are normally consolidated at the testing phase of the application itself. Let us now see how the testing of this integration of the application is going to be done. The testing's commonly done at each module as well as complete application. Let us see how this module level testing is done and then on complete application.

**7.2 Change management - Biometric authentication Methodology**

In the change management of this biometric authentication methodology there are few possible that can be made either in biometric technique or the encryption flow in the application which will be carried out phase by phase after discussing the test result with the change advisory board and other management team before deploying it in the Live operation. There can be an emergency that can be brought about if it required if the hackers if finds a possible approach to reach the confidential information. In that there is going to be a decision going to be made by the emergency change advisory board to deploy the approach immediately here it involves a lot of risk and the downtime requirement if any. So all the information needs to be tested initially itself and should be submitted at the time of need to the management to understand about that approach .So the change requires a proper documentation on the location of the application where the  modifications are made. Once this documentation is done along with the test results, then it provides complete test result with justification for bringing about the change when required for the application. So let see how the process flow is designed for this methodology. This is the process by which the change is going to take place for this biometric authentication methodology. When it is accepted then the It team has analyze on the reason why this change was not accepted it is going to affect the live operations of other application or bringing about this change is not going to bring any effect on the hacker penetrating the network.

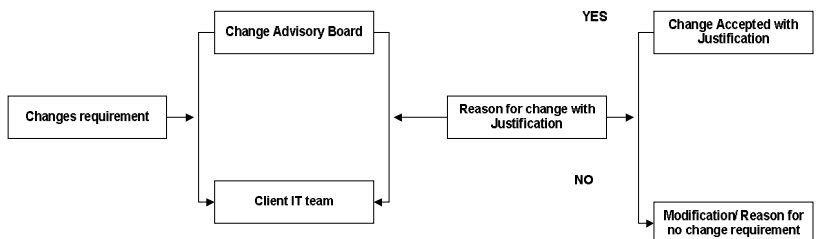

Fig 7 Change management process flow (Biometric authentication methodology)

As far as the biometric authentication is concerned the entry into the server is highly impossible by making these changes the security on the server infrastructure will be high so that the data's that are stored on the server are highly secured without any flaw that allows a third party person to access and view any data's that are stored on the server. Making periodical changes with proper testing on Test server by giving sample inputs will never lead to any issues in the change of an application on a live environment. That is the effective approach for the change management in common. Once this changes are made this needs to document is secured location as these things need not be shared to anyone other the users of this methodology. Bringing out this methodology in the live is not good as it will create an alarm to the hacker that the biometric is going to be used for the authentication purpose. This biometric authentication is going to have a change only to make sure that the process flow is updated periodically with a new one so that there is security that is maintained on the data's that are stored on the server. Once the server security is properly updated then the output can be seen in the security at the enterprise level. Why the client is involved in this change which is going to be a minor change. It is very important to convey the changes that are being made on the server where they are storing their data's or hosting their websites. As this allow them to give their point of view on this change. Then based upon the approvals from both of them will allow the IT team to go ahead with the modification or else suggest them another approach or reason for not making the change. Based upon which the IT team can provide their justification why this change is made and what is the benefit of it behind it. Then accordingly the changes can be made on the authentication methodology.

**7.3 Risk Management - Biometric authentication Methodology**

The risk that is involved in this methodology is very less and that is only the report generation when the change or modification that is done on the process flow. Even this can be avoided when it is tested during the testing phase using the sample inputs. So the risk involved in the biometric image is also an important one that needs to be taken into consideration but that can be justified as a server is accessible by the entire authorized administrator as it is not that a user when registered on a server can access only that server alone. Let me provide you a screenshot on how it looks when a authorized user tries to access any other server in the need of emergency. Here there is no risk involved as this option server name is asked when the biometric image is generated on other server. Let me provide this with an example If the user X has registered his authentication on the server CNHDLADS01 now due to some emergency in restoring the e-mail he access CHNDLML01 then it will ask for this server name along with the biometric image in order to check for the biometric image, the password and the RSASecure ID certification code which is already stored on other server then it will generate the encrypted value to authenticate the user to access the mail server. This process doesn't involve any risk if the testing and the integration are done properly by the development team. Once this is done the user can access the server and restore the e-mails to the user. There are some of the risks that can be handled without any issues in this methodology. This is the main speciality of this methodology as it involves a risk free approach. Even the small risk also can be handled within no time. So the management will get a justifiable reason for this methodology. This is how the risk management is carried out for this methodology. When the justification reason for the risk is not agreed then it needs to be analyzed and produced with a sample input value that is going to convey the management that why this risk is there and how this can be overcome.

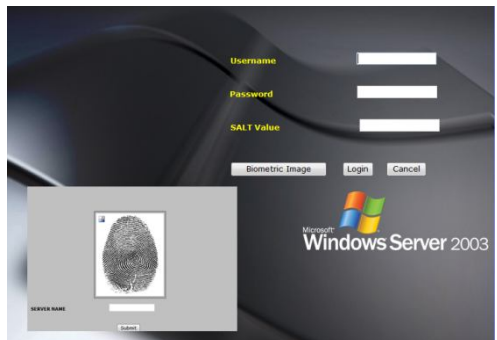

Fig 8 Server authentication (Accessed by other authorized admin)

When a change is made the reports generated on the entire server and the consolidated report generated on the repository server all should get properly aligned with the change that is made in the process flow of the authentication. That can be tested in the testing phase itself so there is only a risk about 0.0000000001% in this approach. So then this it will tell how easy it is for executing and maintaining this approach of server level authentication. The risk management is important as without understanding the risk involved in a application the redundancy can't be developed for an application which is going to be integrated with a Live server.

**7.4 Redundancy Management - Biometric authentication Methodology**

In the redundancy management we are going to see how the biometric images from all the servers are stored on common repository. When there is any fault in a server and it is being reinstalled or replaced then the same biometric image will be loaded back with the same user password. The only change will be the SALT value as it is being generated every 60 seconds RSASecure ID device. How this is going to be carried out and the reason for which it is carried out is to avoid more downtime. Once the server is up and running, the authorized user has to just start login into the server and also to make the process easy as in the live operation this is what they expect from the vendor organization which is maintaining their information. Less downtime with high security is what should be the goal of a data centers at enterprise level. The redundancy of this application is only the above mentioned things as rest all is just deployed if there is any crash in the server or server is being replaced. So what is the situation regarding the log reports, they are taken backup regularly from all the servers on a daily basis and they are sent to all the management team. So there will be no loss in any of the information that is being generated by this authentication methodology. So this is going to tell how redundant the application at the time of emergency is. This will help an organization to keep itself secured with a much easier approach and maintain the same high level of security. This is how all the three parameters is going to keep the organization secured, and also provides a proper approach maintain the changes and handle the situation which are mentioned as risk. In the methodology all parameters are simple and can be restored easily. The important factors that need to be seen in order to achieve this perfection in the implementation of this methodology are training the security policies that need to be set for this methodology. Let us some information's on them after the analysis of this methodology.

## 8. Advantage of this Biometric Authentication Technique

There are lot of advantages of this biometric technique which are as follows 1. These kinds of biometric authentication technique on the server side have not been implemented as the operating systems like Linux and Solaris are considered to be highly security. Even then the hackers are able to hack the data's from the server by breaking up the password. Here this biometric authentication technique will be effective as generating the biometric image is not possible other the authorized IT administrator of that server 2. This biometric authentication has been designed in such a way that it includes a highly secured authentication login technique with encryption algorithm which uses a different approach for generating the final encrypted template using the rearrangement of bits methodology. This makes a very highly secured approach in accessing a server. This kind of authentication can't be seen in recent server authentication. The server authentication application is designed in such a way that it is used in multiple platform which just a small package of deployment to integrate this with the existing authentication technique 3. This authentication technique has a unique feature in authentication when a user is authorized on the server CHNMCADS01 and he access the information on server CHNMCMXL01 then it will ask for the server name where he is registered as it will map with that server for the biometric image and password in order to generate the final encryption template to authenticate the IT administrator. This is the greatest advantage of this approach as the IT administrator need not have to create a new profile on this server in the Emergency Access Mode in order to access this server or need not have to call the respective authorized administrator to access this server. This is the main advantage of this technique which is not in of the current server infrastructure 4. This kind of authentication technique is unique in both the report generation and Emergency access mode as the report is generated and sent automatically sent to the management with the information of the unauthorized user access with his information of the IP address. This authentication technique acts more similar to the network monitoring tool. The Emergency access mode is designed with a lot of limitation with just opening the application which can be used to reset the biometric image/password or can add a new user profile. Nothing else can be seen or can be accessed with this Emergency access. There will be a separate password which will be there with the authorized user of that server. These password should be used anywhere as per the security policy of this application. These Emergency access mode passwords are generated by the provider and provided to the authorized user at the time of delivery of the application. This is accessed only when it is needed with the approvals from the management and after the working hours/non peak hours 5. This methodology has it special encryption technique which has the process of rearrangement of bits which will unique as the output is a number so the final encrypted template will be a number so the hacker even gets this number he will be not be able to get the password, SALT value which keeps changing every login and the biometric image which is unique with all the users. So the authentication is based on biometric image and the user password but the SALT value is to manipulate the final template periodically after every login. This will never give to the hacker on the process flow of this encryption algorithm.

## 9. Future Enhancement of this Biometric Authentication Technique

The future enhancement of this methodology is the file transfer authentication using the biometric and the SALT value. Here these two concepts will be integrated with the file that is being transmitted over the network. There will be authorized user only they can decrypt all the confidential files with their biometric image so the hacker will not be able to read any information without this SALT vale and biometric as the process flow will be something which is used for the server authentication technique. Here the File sent over the network will be encrypted using this application which will be electrically signed using this Biometric image of the authorized user with his password and SALT value and it will be decrypted by the authorized user at the other end. This concept is not related to stenography where the information's are embedded in a common images which was used for the 9/11 world trade center attack. Let us see the process flow of this methodology which will does not includes the encryption process as it is yet to designed for this. As you can see from the diagram below how the authentication for the file transfer has been done. The hackers can decrypt any format even if it 0's and 1's. There are tools that can try to give them the clue on those information's. Even after that how this biometric is going to play a vital role in this authentication process of file transfer. Biometric authentication is something unique as it can't by any other person other than the authorized user. When the biometric authentication is considered for this information security it is sometimes considered not a feasible approach as everything file that is sent has to be encrypted and sent manually by an authorized user. But on the other hand confidential data's when transferred with a proper security authentication technique then it is going to provide high level of security not only for their data's but preventing the hackers from stealing the information's of an organization. This is the authentication technique on which I am going to work on with a new encryption algorithm that will make this encryption process much feasible and much suitable for the enterprise organizations. The complete analysis is going to do with the security attacks that are going on currently. Along with this I am going to work on wider scope of this technique even in ATM transactions and net banking where quite a higher level of security is required. The biometric usage has been bought in Yahoo mail but don't how far this technique is followed by the user as her also the feasibility and the awareness of this biometric usage has to be explained more over the users with laptop will have this biometric option that too on higher configuration models alone.

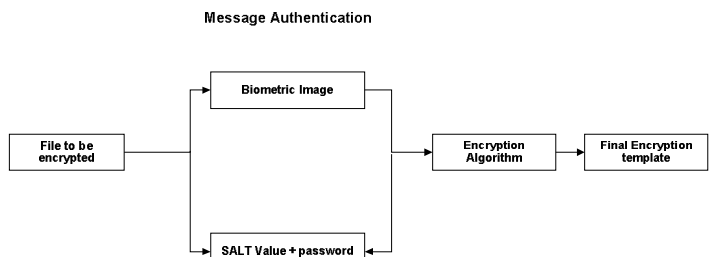

Fig 9 Message authentication using the Biometric and SALT Value

Users who have desktop have to get biometric device as separate component Right now the biometric should be integrated with the existing keyboard so that both can be used for the authentication purpose. Net banking have an image option which needs to be selected while using the net banking for the first time. Then it will be displayed for the user when he accesses the account for the consecutive time. But it will not considered as a proper authentication technique as we hear a lot of hackers who are stealing customer's information even though the web site is secured by a third party security provider. That is where the biometric integration with the information is going to play a vital role as when the information is encrypted with the image then dividing them is not an easy task when compared to the information just encrypted with the encryption algorithm. That is where we can see the real usage of biometric integration with confidential information. The biometric approach of authenticating the user is considered to be the most positive sign regarding the biometric techniques in today world where requirement of security is high. When the biometric is used with any form of security section say authenticating a resident person, employee of an organization, authorized user of a server (In infrastructure support) it has been feasible, reliable and above all the security that it provide is very high as each biometric image that is generated is unique. But here comes the technique that can be used and it varies with the sector like Retina and face recognition can be for employee authentication, retina and finger print for laptop authentication. Based upon these criteria's, going to design the next authentication technique for messaging system at enterprise organizations. This application is going to maintain a high level of security for the messages that are been shared between the clients and the branches of the organization. This application is going to maintain the logs in similar fashion but the server utilized with a centralized with a backup server as a source of redundancy. This application is going to be operating system based which will be implemented at the enterprise and not at a consumer level.

## 10. Conclusion

In this chapter we have seen how a biometric integration is going to used in Server authentication at enterprise level with the SALT value and encryption algorithms. When the biometric technique is used in authentication what are the parameters that needs to be followed and how it needs to taken into consideration with the management views are been discussed in this chapter. Then finally we have seen how to provide training for this application as training is considered as an important part of the IT transition. Based upon the assessment only the enhancement of this application also can be carried. But this will discussed in the initial stage itself before providing this application as when the security norms are signed and followed it should be followed like a holy book. As the IT admin should be simultaneously updated with the recent security threat and what is the solution that is enhanced from the application side. Then we have seen about the testing phase of the application during the implementation of the application at enterprise level and the information's that need to be checked while implementation of this authentication technique. Then we have seen recommended technique in biometric, which are completely based on the above parameter based on both management and the biometric technique parameters. Finally it is all about the IT security policy which is set for the application based on the current policy norms that are set by ISO standards, information policy as mentioned in CISSP, CISA and CISM. All these deal with the information security policy. These are some of topic we have discussed with the some real time examples of biometric authentication in other technologies. Then we have a topic that is "Need to know principle" which is a wonderful topic that tell about the limitations that needs to be set in security policies. These are the information's that has been discussed in this chapter. It is like providing a complete overview of this Biometric integration with the SALT value at the enterprise level server authentications. Here it not only show the technique but followed by analysis, view of management with the important parameter and how this technique is going to be better that just a normal password authentication.

## 14. Acknowledgement

First and foremost I would like to thank almighty for giving me the courage, confidence and the strength to do this paper with a lot of dedication and complete this paper with all the possible analysis that was required for this topic. Then I would thank my parents who have always been my support in carrying any task that were related both my job and academics. They have been my role model right my childhood day so a special thanks goes to them as well as I have their blessing for publishing this chapter successfully. Al last I would like to the entire Intech open access publisher for keeping me updated on the days left for my work to the updation in the website that is in the user account updation. My heart full thanks go for the entire team of Intech open access publisher who made my journey smooth for publishing this full chapter. Finally I would be happy to publish this chapter for all those innovators who are eager to know more about biometric usage in enterprise level authentication techniques. This chapter will be great help and useful information provider for those who are working in the information security, infrastructure support and implementation team etc.